\documentclass[]{article}
\usepackage{graphicx}
\usepackage{amsmath}
\usepackage{hyperref}
\usepackage{authblk}
\usepackage{color}
\usepackage[top=3cm, bottom=2cm, outer=3cm, inner=3cm]{geometry}

\begin{document}
	\title{On The Sound Speed in Hyperonic Stars}
	
	\author[$a\,\star$]{T.~F.~Motta}
	\author[$b$]{P.~A.~M.~Guichon}
	\author[$a$]{A.~W.~Thomas}
	
	\affil[$a$]{CSSM and Department of Physics, University of Adelaide, SA 5005 Australia}
	\affil[$b$]{IRFU-CEA, Universit\'{e} Paris-Saclay, F91191 Gif sur Yvette, France}
	\graphicspath{{./}}
	
	\maketitle
	\begin{abstract}
		We build upon the remarkable, model independent constraints on the equation of state of dense baryonic matter established recently by Annala {\em et al.}~\cite{kurkelanature}. Using the quark-meson coupling model, an approach to nuclear structure based upon the self-consistent adjustment of hadron structure to the local meson fields, we show that, once hyperons are allowed to appear in dense matter in $\beta$-equilibrium, the equation of state is consistent with those constraints. As a result, while one cannot rule out the occurence of quark matter in the cores of massive neutron stars, the available constraints are also compatible with the presence of hyperons.
	\end{abstract}

	\section{Introduction}
	At the present time there is a great deal of interest in the existence of very heavy neutron stars (NS), with masses of order 2M$_\odot$ or larger. Given the extreme conditions at the core of such stars, with densities much greater than the density of nuclear matter ($n_0$), many authors have suggested that they may contain deconfined quark matter -- see 
Refs.~\cite{Baym,qmrevBastian:2018wfl,qmrevBombaci:2016unk,qmrevJaikumar:2006rh} for recent reviews.
	
	Amongst the most important achievements in experimentally constraining the equation of state (EoS) of neutron stars was the measurement of the mass of a pulsar equal to 1.97$\pm0.04$M$_\odot$ \cite{Demorest}, which was later revised to 1.908$\pm0.016$M$_\odot$ \cite{Arzoumanian}. This discovery and the later measurement by Antoniadis {\em et al.}~\cite{Antoniadis:2013pzd} and Cromartie {\em et al.} \cite{Cromartie} provided important new constraints. One, of an instrumental nature, was that models that do not lead to stars with masses at least as large as 1.9M$_\odot$ are incorrect and should be abandoned. Another, more of an epistemic nature, was based on the following puzzle. A star this massive \textit{must} have an innermost core that is incredibly dense; most models point, at the very least, to 5 times nuclear matter density. However, at such densities the argument for a fundamental change in the nature of matter becomes compelling. Indeed, it is possible that neutrons and protons might dissolve and the relevant degrees of freedom become quarks and gluons; that is, the core might consist of quark matter (QM). 
	
	An alternative, to which rather less attention has been paid, is that  perhaps the Fermi momentum and chemical potential of the nucleons are so high that hyperons can be created, by which point, because of the Pauli blocking, they would be stable (see e.g. \cite{2018,hypGomes:2017zkc,hypMalfatti:2020onm,hypAdhitya:2020nlq,Weise:2019mou}). This possibility should also be of enormous interest as strangeness is now considered to be one of the new frontiers of nuclear 
physics~\cite{Nagae:2010zza,Thomas:2018hux,Beane:2012ey}. While rare ion facilities push the boundaries of the known atomic nuclei in $N$ and $Z$, strangeness may be thought of as a third axis along which much less is known. In the strangeness minus one ($S=-1$) sector quite a few 
$\Lambda$-hypernuclei have been measured and we know that almost no $\Sigma$-hypernuclei  are bound~\cite{Tamura:2019yxd,Gal:2004cf}. However, when it comes to the $S=-2$ cascade hyperon (or $\Xi$, where two light quarks in a nucleon have been replaced by strange quarks), so far only one hypernucleus has been discovered~\cite{Nagae:2019uzt}. For many years NS modeling including hyperons led to maximum masses of order 1.6M$_\odot$, which was far too low. Only with the inclusion of three-body forces~\cite{2007}, was it found to be possible to generate NS at 2M$_\odot$ with hyperons. 
	
	In a stimulating recent study, Annala {\em et al.}~\cite{kurkelanature} used a novel interpolation method based upon the speed of sound in-medium, $c_s$, to generate a large, model independent set of equations of state (EoS). These EoS were required to be consistent with the existence of NS up to 1.97 M$_\odot$, as well as the constraints on tidal deformability deduced from the neutron star merger data obtained from GW170817~\cite{gw}. By identifying regions within massive NS with $\gamma$, the polytropic index, below 1.75 (conversely $c_s^2 < \frac{1}{3}$) and realizing that these were inconsistent with an EoS involving only nucleons\footnote{They state, in Ref.~\cite{kurkelanature}, that ``both CET calculations and hadronic models generically predict $\gamma \approx 2.5$ above saturation dens."}, it was deduced that the existence of quark matter cores in massive NS should be considered the standard scenario.
	
	Here we explore the consequences for the speed of sound and the polytropic index of including hyperons and imposing $\beta$-equilibrium on the dense matter in the core of massive NS. As noted by a number of authors, starting with Stone {\em et al.}~\cite{2007}, this naturally implies that NS with masses above $\sim$ 1.6M$_\odot$ must contain hyperons. We find that the softening of the EoS as each new species of hyperon appears lowers the speed of sound below $\frac{1}{3}$ and the polytropic index well below 1.75. Thus, while one cannot rule out the appearance of quark matter in the cores of such massive stars, the model independent analysis of Annala {\em et al.} and the current observational constraints on NS properties are both also consistent with the appearance of hyperons.
	
	The structure of this work is as follows: in the next section we review the claims and methodology of Ref.~\cite{kurkelanature}; in section~\ref{sec:QMC} we briefly review the basic assumptions of the QMC model, and finally, in section~\ref{sec:compare}, we compare the two results before proceeding to the conclusions.
	
	\section{Methods}
	\subsection{Model-Independent Analysis}
	\label{sec:naturepaper}
	In Ref.~\cite{kurkelanature} the authors aimed to provide a model independent analysis of the relation between the shape of the EoS of dense matter and the experimental constraints on NS properties. They randomly generated EoS that were selected to respect key high and low density limits, namely the perturbative, conformal limit of QCD at high density and the low density limit given by chiral effective field theory (CET). Acceptable EoSs within this set were then required to satisfy important astrophysical constraints, such as an upper limit on the maximum mass of a NS of at least 1.97M$_\odot$ and the tidal deformability limits from GW170817~\cite{gw}.
	
	The novel, model independent method that was employed to generate such equations of state was based upon multiple different interpolations which were compared with each other, namely
	\begin{enumerate}
		\item a piecewise polytropic form for the pressure, $p_i=k_in^{\Gamma_i}$,
		\item an interpolation of the adiabatic index itself, $\Gamma(p)$, through Chebyshev polynomials,
		\item a piece-wise interpolation of the speed of sound in terms of linear functions of the baryon chemical potential, $c_s^2(\mu)$.
	\end{enumerate}

	The EOS generated in this manner were studied and categorised via the following methodology. A polytropic  index, $\gamma={d\ln p}/{d \ln \varepsilon}$, within a certain range, namely $\gamma > 2$ was identified as being characteristic of hadronic matter. Conversely, the smallest density after which $\gamma<1.75$ was taken to identify a region of quark matter.
	
	On the basis of the large family of EoS generated in this model independent way, Annala {\em et al.}~\cite{kurkelanature} found that, although stars of the hadronic type {\em containing nucleons only}, according to the categorisation discussed above, do reproduce well the properties of stars with moderate mass (roughly $1.5$M$_\odot$), the highest mass stars are reproduced overwhelmingly by EOS that show evidence of quark matter. For this reason they concluded that one should consider the existence of quark matter in the cores of massive NS as the standard scenario.
	
	\subsection{Equation of State with Hyperons}\label{sec:QMC}
	The quark-meson coupling (QMC) model builds a description of nuclear matter and finite nuclei based upon the self-consistent modification of the structure of the bound baryons in the relativistic mean fields generated by meson coupling to the confined quarks~\cite{Guichon:1987jp,Guichon:1995ue,Stone:2016qmi,Saito:2005rv,Krein:1998vc}. The internal structure of the baryons is usually described by the MIT Bag model~\cite{bag}, although the NJL model has also been employed~\cite{Bentz:2001vc,Whittenbury:2015ziz}. The quarks confined inside their bag interact with quarks in neighbouring baryons via the exchange of meson fields. We include one-gluon-exchange (OGE) between the quarks in the bag and obtain an expression for the baryon effective mass 
	\begin{eqnarray}
	\label{massstar}
	M_{N}^\star=\frac{\Omega_{u} N_{u}+\Omega_{d} N_{d}+\Omega_{s} N_{s} -z_0}{R_{B}}+\mathcal{B} V_{B} + \Delta E_{M}
	\end{eqnarray}
	where $\Omega_q/R_B$ is the energy eigenvalue for the quark $q$ and the OGE hyperfine color interaction contribution is $\Delta E_M$. Both of these depend non-linearly on the scalar meson mean field, 
	$\bar\sigma$~\cite{Guichon:2008zz}. The bag pressure, zero point fluctuations $z_0$ and strength of the gluon exchange, given in terms of the strong coupling $\alpha_c$, are all fitted to reproduce the masses of the entire baryon octet in free space. The resulting expression can be fitted to an analytic form such as
	\begin{eqnarray}
	\begin{aligned}
		M^\star_B(\bar\sigma) &= M_B - g_{\sigma B} (\bar\sigma)\bar\sigma\\
		&=M_B -g_{\sigma B} \bar\sigma + \frac{d_B}{2}(g_{\sigma B} \bar\sigma)^2	\, ,
	\end{aligned}
	\end{eqnarray}
	where $d_B$ is the scalar polarizability of baryon $B$.
	
	The procedure just outlined yields a density dependent mass for every baryon in the octet without adding~\textit{any} new parameters. This density dependence is equivalent to the inclusion of many-body forces, as the interaction between a pair of baryons is modified by the presence of others~\cite{Guichon:2006er,Guichon:2004xg}. The most important effects are related to the effective three-body forces, which are repulsive, depend on the particular baryons and involve no new parameters. The automatic generation of these forces was the reason that the QMC model was able to predict NS with masses of order 2 M$_\odot$, even when hyperons were included, before any such stars had been found~\cite{2007}.
	
	The free space baryon-meson couplings are fitted to reproduce standard nuclear matter parameters (as Ref.~\cite{motta2019}), namely, the saturation density, symmetry energy and binding energy per nucleon, respectively
	\begin{eqnarray}
	&n_0=0.148\text{fm}^{-3},\\ &S=30\text{MeV},\\ &\mathcal{E}=-15.8\text{MeV} \, .
	\end{eqnarray}
	Taking the vector meson masses to have their experimental values~\cite{pdg}, there remain just two parameters to be  fixed, namely $R_B$ and the mass of the sigma meson,  $m_\sigma$. We choose these to be, respectively, $0.8$fm (results with different bag radii have been published \cite{2018,kay}, however, they do not alter any of the conclusions drawn here) and $m_\sigma=700$MeV.
	
	The detailed expressions for the energy density are shown in the Supplementary Material. For the intents and purposes of this discussion, suffice it to say that in minimising that energy density with respect to $\beta$-equilibrium conditions we do find that hyperons must be present in the cores of neutron stars with masses above 1.8M$_\odot$. 
	%
	
	\section{Neutron Stars with hyperons}
	\label{sec:compare}
\begin{figure}[t]
		\centering
		\includegraphics[width=0.9\textwidth]{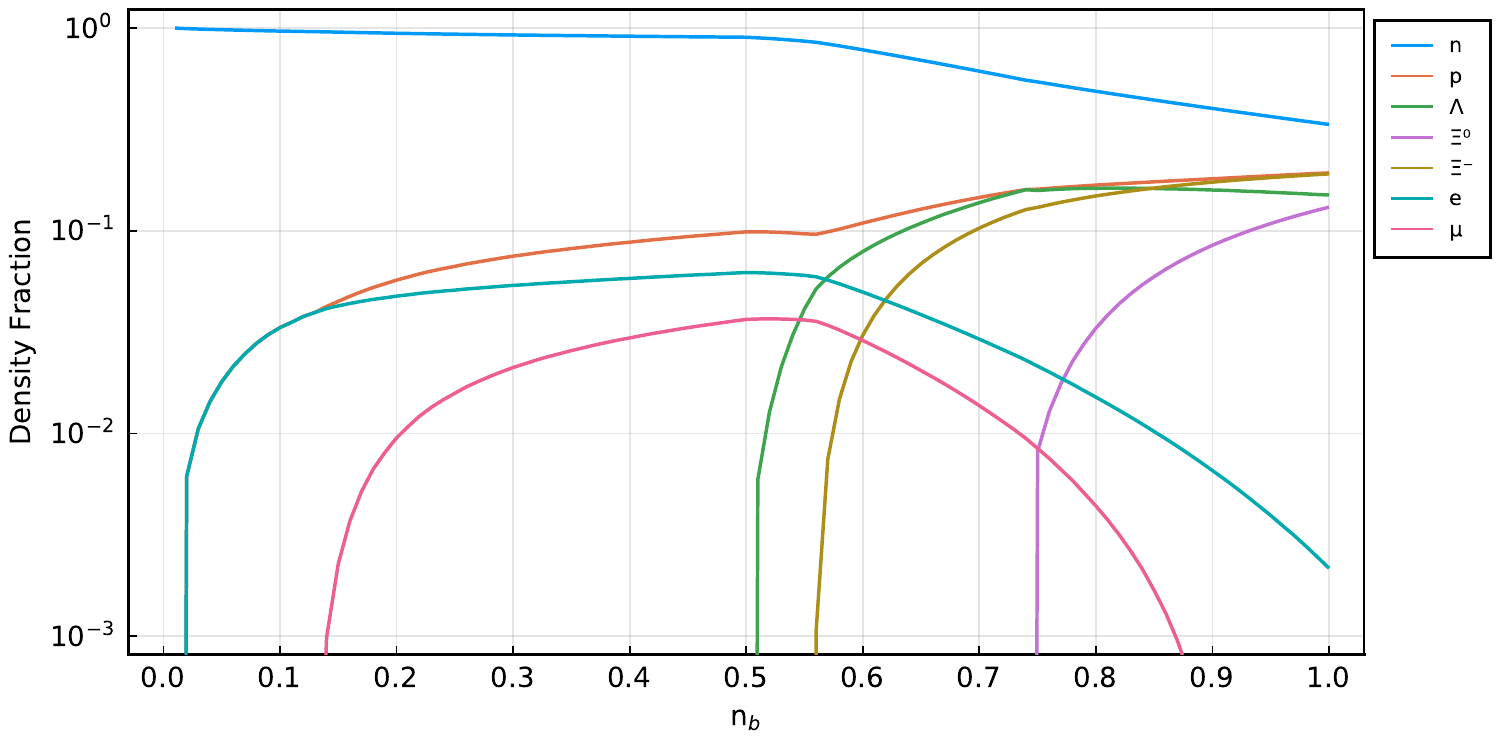}
		\caption{Fraction of number density per baryon density of each baryon species as a function of the total baryon density.}\label{fig:species}
	\end{figure}

The predictions of the model outlined above for NS properties were fully explained in 
Ref.~\cite{motta2019}. In particular, the maximum mass including hyperons
was of order 1.95 M$_\odot$, which is compatible with J1614, which has smaller error, within one standard deviation and with other measurements of heavy stars within two standard deviations (see next section).
The radius and tidal deformability of a NS of mass 1.4 M$_\odot$ were 12.8km and 420, respectively, consistent with the constraints deduced from the GW170817 data and used in 
Ref.~\cite{kurkelanature}. Figure~\ref{fig:species} shows the fraction of each species as a function of the baryon density, noting that the largest mass stars in our model have core densities of roughly 1fm$^{-3}$.

\begin{figure}
		\centering
		\includegraphics[width=0.9\textwidth]{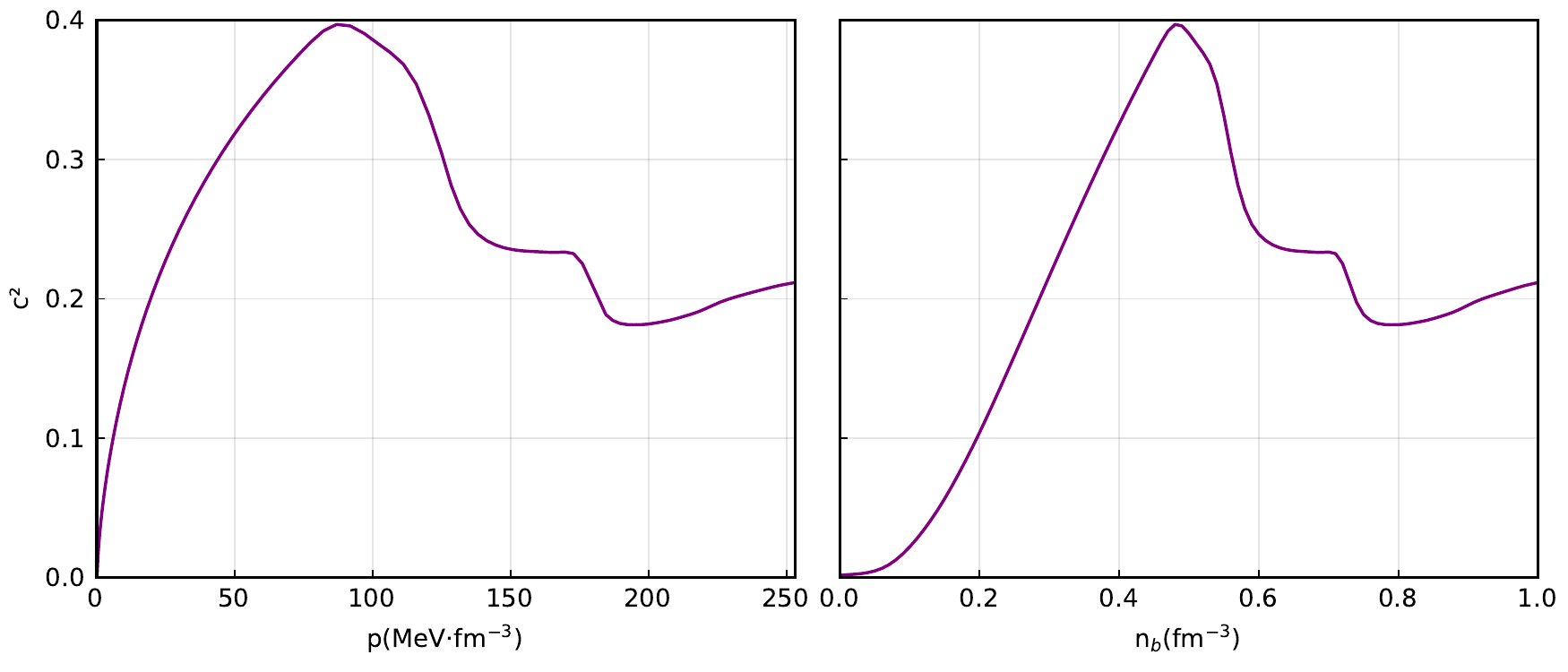}
		\caption{The speed of sound squared, $c_s^2$, is illustrated as a function of pressure and baryon number density.}
		\label{fig:cs2}
	\end{figure}

As shown in Fig~\ref{fig:cs2}, the EoS in the QMC model exhibits a sudden decrease in the speed of sound as the density increases above 0.5 fm$^{-3}$. That decrease, which in other models is indeed characteristic of the introduction of a QM phase, here is a natural consequence of passing the threshold at which hyperons start to be generated. As can be seen by comparing Figs.~\ref{fig:species} and \ref{fig:cs2}, the fairly dramatic decreases in the speed of sound coincide with the appearance of first the $\Lambda$, then 
	the $\Xi^-$ and subsequently the $\Xi^0$. At each threshold the appearance of the hyperon softens the EoS, because at threshold one is adding energy but little momentum, and this in turn reduces the speed of sound and the adiabatic index.

\begin{figure}
		\centering
		\includegraphics[width=0.9\textwidth]{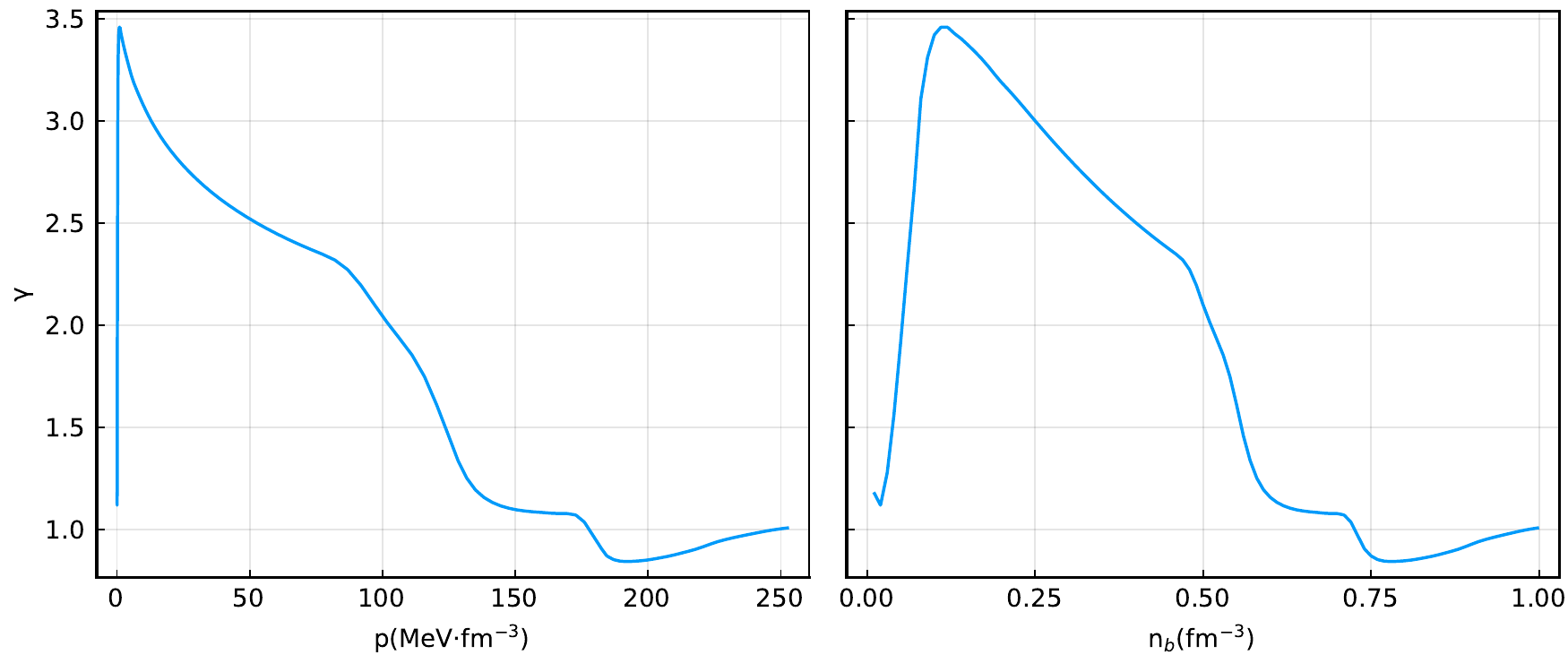}
		\caption{We show the polytropic index, $\gamma = d(\ln P)/d(\ln \epsilon)$, as a function of pressure and baryon number density.}
		\label{fig:gamma}
	\end{figure}

	As can be seen in Fig.~\ref{fig:gamma}, the value of the adiabatic index also shows a sudden decrease when the density reaches the first hyperon threshold and again later  with the introduction of the $\Xi^0$. It is especially important to note that the natural appearance of hyperons in the QMC model leads to values of the polytropic index, $\gamma$, well below $1.75$. 

\begin{figure}[t]
		\centering
		\includegraphics[width=\textwidth]{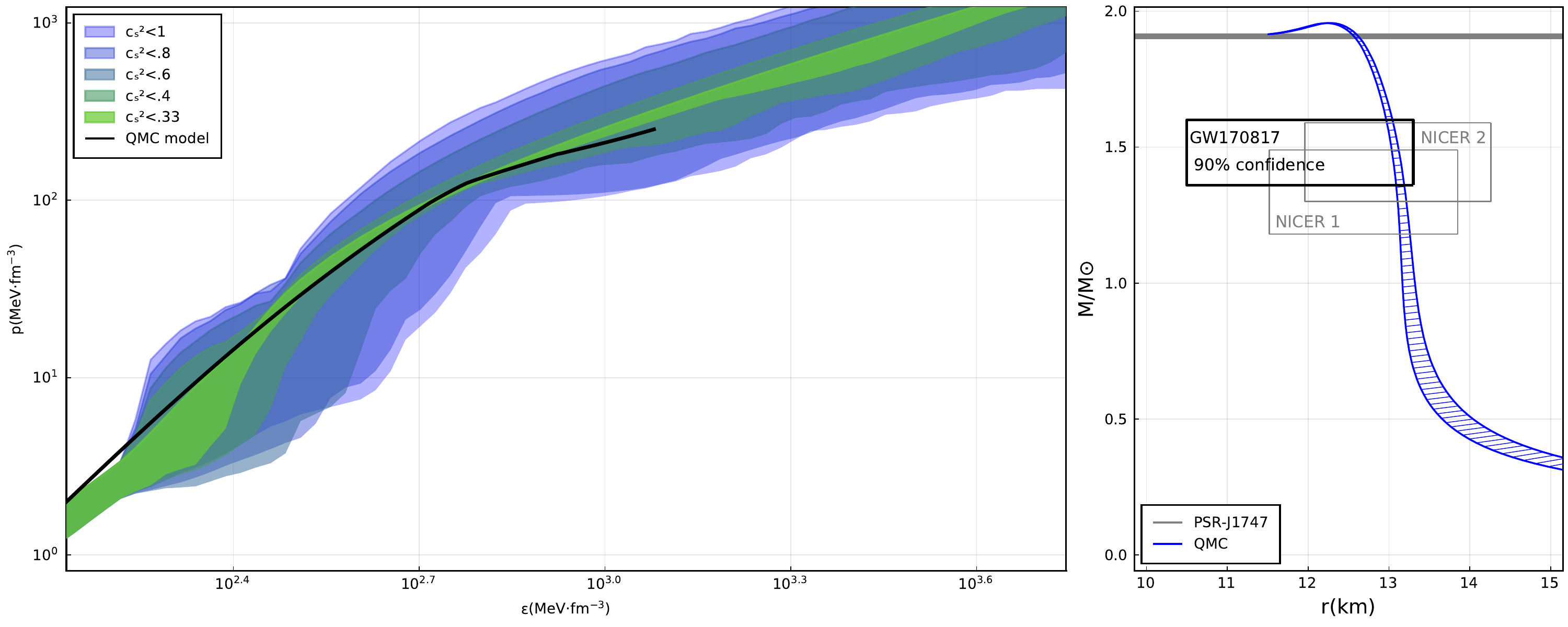}
		\caption{Comparison of the EoS in the QMC model, including hyperons as required by $\beta$-equilibrium, with the model independent constraints found in Ref.~\cite{kurkelanature}. The EoS labelled QMC on the left corresponds to the mass-radius diagram on the right. The model also agrees with the constraints on GW170817 \cite{gw} and agrees with Ref.~\cite{Nattila} within one or two standard deviations, although Ref.~\cite{Nattila} is still somewhat model dependent and not yet sufficiently strong to exclude models. The bands in radius from the NICER experiment labelled NICER1 and NICER2 correspond respectively to \cite{Nicer1Riley_2019,Nicer2Miller_2019}. The blue hatching represents the systematic error due to the joining of the crust EoS, which we use the one described in Ref.~\cite{DD2Y}, with the core EoS.}
		\label{fig:bands}
	\end{figure}

	Finally, Fig.~\ref{fig:bands} shows the EOS predicted within the QMC model overlaid on the bands generated using the model independent techniques presented in Ref.~\cite{kurkelanature}. Clearly the present model is completely consistent with these general constraints\footnote{Note that, in Ref.~\cite{kurkelanature}, the low density side of the equation of state consists of matter in $\beta$-equilibrium, taken from Ref.~\cite{Hebeler:2013nza}. For consistency, we show here our equation of state of infinite nuclear matter in beta equilibrium at all densities, and do {\em not} replace the low density region with a realistic crust EoS.}. However, as we have emphasised, the QMC model EoS contains no quark matter but rather hyperons.
	%
	
	\section{Conclusion}
	The model independent constraints on the EoS of dense matter deduced in Ref.~\cite{kurkelanature} are of great interest. It is remarkable that the kink in the plot of pressure versus energy density, at a baryon density of order 0.5fm$^{-3}$, coincides with that found when hyperons are included. This is particularly interesting given that the appearance of hyperons played no role in the work of Ref.~\cite{kurkelanature}. Rather, the connection to perturbative QCD at very large densities was key to the appearance of the kink in that work. 
	
	While acknowledging the power of the techniques employed by Annala {\em et al.}~\cite{kurkelanature}, we have presented evidence for a different interpretation. Using the QMC model, which not only produces a very good description of the properties of finite nuclei across the periodic table but is consistent with all observational constraints relating to NS, we find that the kink in the EoS and the consequent reduction in the speed of sound and the polytropic index, $\gamma$, is associated with the appearance of hyperons in the dense matter which is in $\beta$-equilibrium. 
	
	In the present work the polytropic index does indeed fall below 1.75 for energy densities above 600 MeV/fm$^3$. However, this is a signal of the appearance of hyperons rather than quark matter. Of course, the current calculations cannot exclude the possibility that stars with masses beyond about 1.8M$_\odot$ might contain quark matter. Nevertheless, one equally cannot exclude that the cores of such stars consist instead of a significant fraction of hyperons. Clearly, it is a fascinating challenge for future work to find new ways to test which of these two possibilities is chosen in Nature.
	
	\section*{Acknowledgement}
	This work was supported by the University of Adelaide and by the Australian Research Council through Discovery Project DP180100497.
	\section*{Author Contributions}
	All authors participated in editing the manuscript as well as in the derivation of the current version of the model. The numerical calculations were implemented by T.M.
	
	\section*{Competing Interests statement}
	The authors declare no competing interests.
	
	
\end{document}